\documentclass[12pt]{article}

\begin{document}
\begin{center}
{\bf  CHARGE RADII AND MAGNETIC POLARIZABILITIES \\
OF $\rho$ and $K^*$ MESONS IN QCD STRING THEORY} \\
\vspace{5mm}
 S. I. Kruglov \\
\vspace{5mm} \textit{International Educational Centre, 2727
Steeles Ave. West, Suite 202, \\Toronto, Ontario, Canada M3J 3G9}
\end{center}

\begin{abstract}
The effective action for light mesons in the external uniform
static electromagnetic fields was obtained on the basis of QCD
string theory. We imply that in the presence of light quarks the
area law of the Wilson loop integral is valid. The approximation
of the Nambu-Goto straight-line string is used to simplify the
problem. The Coulomb-like short-range contribution which goes from
one-gluon exchange is also neglected. We do not take into account
spin-orbital and spin-spin interactions of quarks and observe the
$ \rho $ and $K^*$ mesons. The wave function of the meson ground
state is the Airy function. Using the virial theorem we estimate
the mean charge radii of mesons in terms of the string tension and
the Airy function zero. On the basis of the perturbative theory,
in the small external magnetic field we find the diamagnetic
polarizabilities of $\rho $ and $K^*$ mesons: $ \beta_\rho
=-0.8\times 10^{-4}~\mbox{fm}^3$, $\beta _{K^*}=-0.57\times
10^{-4}~\mbox{fm}^3$
\end{abstract}

\section{Introduction}
One of the important problems of particle physics is the
confinement of quarks. There is progress in understanding of
properties of mesons as a system in which quarks and antiquarks
are connected by the relativistic string with a Nambu-Goto
self-interaction [1]. This binding interaction becomes strong at
large distances and therefore it is impossible to describe the
phenomenon using the perturbative approach. There are some
difficulties in evaluating meson characteristics in the general
case of a complicated string configuration. Naturally, as a first
step, we make some approximations and model assumptions to
simplify the calculations. So here we consider the straight-line
string as a simple configuration and quarks attached to the ends
of the string. Such configurations were studied in [2]. In the
present paper we investigate mesons in external, constant, and
uniform electromagnetic fields and use the path integral approach.
It should be noted that in potential-like models [3-5] meson
characteristics are described reasonably. But in these approaches
there are some assumptions: (i) the relativistic invariance is
only the approximate, and (ii) the constituent quark masses are
used (but not current quark masses).

The recent development of the QCD string approach [6-11] showed
good results in describing heavy quarkonia, baryons and glueballs.
The QCD string theory takes into account the main nonperturbative
effects of strong interactions: chiral symmetry breaking (CSB) and
the confinement of quarks. Chiral symmetry breaking gives a
nonzero quark condensate ($\left\langle \bar qq\right\rangle $).
As a result the light quarks ($u$, $d$ - quarks) with current
masses $m_u\simeq m_d\simeq 7$ MeV acquire the dynamical masses $
\mu _u\simeq \mu _d\simeq 320$ MeV. This phenomenon is important
for light pseudoscalar mesons as they possess the Nambu-Goldstone
nature. To get low masses of pseudoscalar mesons one needs to take
into account spin interactions of quarks. In [8,9] CSB was
explained by the nonvanishing density of quark (quasi) zero modes
in the framework of QCD. Then the familiar PCAC (partial
conservation of axial vector current) theorems and the soft pion
technique are reproduced. There are different stochastic vacuum
configuration which are responsible for CSB: instantons
(anti-instantons), pieces of (anti-)self-dual fields (for example,
torons or randomly distributed lumps of field), and others. The
necessary requirement is to have zero fermion modes. The
condensation of zero modes leads to CSB. The confinement of quarks
does not allow them to be observed; i.e. quarks cannot move
outside of hadrons in large distances relative to each other. This
was confirmed by Monte-Carlo simulations and experiments. Both
nonperturbative effects of strong interactions can be explained by
introducing stochastic gluon vacuum fields with definite
fundamental correlators [6,7]. Then the linear potential between
quarks appears and it provides the confinement of quarks. Besides,
Regge trajectories are asymptotically linear with a universal
slope [7]. So the method of vacuum correlators in nonperturbative
QCD and the dynamics of zero modes give the explanation of the
double nature of light pseudoscalar particles (pions, kaons and
others) as Nambu-Goldstone particles and as the quark-antiquark
system with the confining linear potential. It should be noticed
that confinement prevents the delocalization of zero modes over
the whole volume [8], i.e., stabilizes the phenomena of CSB.

In the present approach we make some assumptions. So we treats
spin degrees of freedom as a perturbation and therefore it is
questionable to apply this scheme to pions and kaons. Only $\rho $
and $K^{*}$ mesons are considered here because the energy shift
for them due to the hyperfine spin interaction is below $100$ MeV
[4]. Here short-range spin-orbital ${\bf L}\cdot {\bf S}$ and
spin-spin ${\bf S_1}\cdot {\bf S_2}$ interactions are not taken
into account. Besides we neglect the Coulomb-like short-range
contribution due to the asymptotic freedom of QCD. This
contribution is important only for heavy quarkonia [5]. We imply
also that in the presence of light quarks the structure of the
vacuum yields an area law of the Wilson loop integral. The
restriction of the leading Regge trajectories is used as we
consider here only $\rho$ and $K^{*}$ mesons.

It is important to calculate different intrinsic characteristics
of hadrons on the basis of QCD string theory and to compare them
with experimental values. It will be the test of this scheme. The
charge radius (and electromagnetic form-factors) and
electromagnetic polarizabilities of mesons are fundamental
constants which characterize the complex structure of particles.
These values for some mesons are known from the experimental data
and therefore the estimation of them is reasonable. So we can
check our notion about the vacuum structure by coinciding
experimental data and theoretical predictions. In [12] we made the
crude estimate of the charge radius and the electric
polarizability of mesons and nucleons. Here we evaluate the
mean-squared radius (using the virial theorem) and the magnetic
polarizability of $\rho$ and $K^*$ mesons.

The electromagnetic polarizabilities of hadrons $\alpha $, $\beta
$ enter the induced electric ${\bf D}=\alpha {\bf E}$ and magnetic
${\bf M} =\beta {\bf H}$ dipole moments, where ${\bf E}$, ${\bf
H}$ are the strengths of electromagnetic fields. As a result there
is a contribution to the polarization potential [13,14] as follows
$$
U(\alpha ,\beta )=-\frac 12\alpha {\bf E}^2-\frac 12\beta {\bf
H}^2. \eqno(1.1)
$$

Electromagnetic polarizabilities are fundamental low-energy
characteristics of strong hadron interactions and therefore they
can be calculated in the framework of non-perturbative quantum
chromodynamics $-$ QCD string theory.

This paper is organized as follows. In Sec. 2 after describing the
general background we derive the effective action for mesons in
external electromagnetic fields. The ground state and charge radii
of particles are found on the basis of exact solutions and the
virial theorem in Sec. 3. Section 4 contains the evaluation of the
diamagnetic polarizabilities of $\rho$ and $K^*$ mesons using the
perturbative expansion in a small magnetic field. In the
conclusion we made a comparison of our results with other
approaches.

Units are chosen such that $\hbar =c=1$.

\section{Effective Action for Light Mesons}

To get the effective action for mesons in an external
electromagnetic field we use the Green function of the quarks
possessing spins in Minkowski space [see Eq. (A8) in the
Appendix)]:
$$
S(x,y)=i\int_0^\infty ds\int_{z(0)=y}^{z(s)=x}Dz\;\left( m_1-\frac
i2\gamma _\mu \dot z_\mu \left( t\right) \right) P\exp \biggl
\{i\int_0^s \biggl [ \frac 14\dot z_\mu ^2(t)-m_1^2
$$
$$
+ e_1\dot z_\mu (t)A_\mu ^{el}(z)+\Sigma _{\mu \nu }\left(
e_1F_{\mu \nu }^{el}+gF_{\mu \nu} \right) \biggr ]dt\biggr \}\Phi
(x,y),\eqno(2.1)
$$

where $z_\mu (t)$ is the path of the quark with the boundary
conditions $ z_\mu (0)=y_\mu $, $z_\mu (s)=x_\mu $ and
$$
\Phi \left( x,y\right)=P\;exp\biggl \{ ig\int_y^xA_\mu dz_\mu
\biggr \} \eqno(2.2)
$$

is the path-ordered product (the parallel transporter), $A_\mu $
is the gluonic field and $g$ is the coupling constant. Neglecting
quark-antiquark vacuum loops and omitting the annihilation graph,
the Green function of mesons (the quark-antiquark system) takes
the form [7]
$$
G(x,\bar x;y,\bar y)=\mbox{tr}\left\langle \gamma _5S(x,y)\Phi
\left( y,\bar y\right) \gamma _5S(\bar y,\bar x)\Phi \left( \bar
x,x\right) \right\rangle, \eqno(2.3)
$$

were the brackets $\left\langle ...\right\rangle $ are the
averaging over the external vacuum gluonic fields with the
standard measure $\exp \left[ iS\left( A\right) \right]$.
Inserting (2.1) into (2.3) we find the expression
$$
G(x,\bar x;y,\bar y)=\mbox{tr}\int_0^\infty ds\int_0^\infty d\bar
s\int_{z(0)=y}^{z(s)=x}Dz\left( m_1+\frac i2\gamma _\mu \dot z_\mu
\left( t\right) \right)
$$
$$
 \times \int_{\bar z(0)=\bar y}^{\bar z(\bar s)=\bar x}D\bar
z\left( m_2-\frac i2\gamma _\mu \dot{\bar z}_\mu \left( \bar
t\right) \right) ~P_\Sigma \exp \biggl \{i\int_0^s\biggl [\frac
14\dot z_\mu ^2(t)-m_1^2+e_1\dot z_\mu (t)A_\mu ^{el}(z)
$$
$$
+e_1\Sigma _{\mu \nu }F_{\mu \nu }^{el}\left( z\right) \biggr ]
dt+i\int_0^{\bar s}\left[ \frac 14\dot{\bar z}_\mu ^2(\bar
t)-m_2^2+e_2\dot{\bar z}_\mu (\bar t)A_\mu ^{el}(\bar z)+e_2\Sigma
_{\mu \nu }F_{\mu \nu }^{el}\left( \bar z\right) \right] d\bar t
\biggr \}
$$
$$
\times\left\langle \exp \left\{ ig\Sigma _{\mu \nu }\left[
\int_0^sF_{\mu \nu }\left( z\right) dt-\int_0^{\bar s}F_{\mu \nu
}\left( \bar z\right) d\bar t\right] \right\} W(C) \right\rangle,
\eqno(2.4)
$$

where $P_{\Sigma}$ is the ordering operator of the spin matrices
$\Sigma _{\mu \nu }$; $e_1$, $e_2$ are charges and $m_1$, $m_2$
are current masses of the quark and antiquark; $z_\mu (t)$, $\bar
z_\mu (\bar t)$ are the paths of the quark and antiquark with the
boundary conditions $z_\mu (0)=y_\mu $, $ z_\mu (s)=x_\mu $, $\bar
z_\mu (0)=\bar y_\mu $, $\bar z_\mu (\bar s)=\bar x_\mu $ and
$\dot z_\mu (t)=\partial z_\mu /\partial t$. Here we used the
properties of $\gamma $ matrices: $\left\{ \gamma _5,\gamma _\mu
\right\} =0$, $\left[ \Sigma _{\mu \nu },\gamma _5\right] =0$. As
compared with [7,17] we added the interaction of charged quarks
with the external electromagnetic fields. The gage - and
Lorenz-invariant Wilson loop operator is given by
$$
W(C)=\frac{\mbox{tr}}{N_C}P\exp\left\{ ig\int_CA_\mu dz_\mu
\right\}, \eqno(2.5)
$$

where $N_C$ is the color number, and $C$ is the closed contour of
lines $x\bar x$ and $y\bar y$ connected by paths $z(t)$, $\bar
z(\bar t)$ of the quark and antiquark. The Wilson operator (2.5)
contains both the perturbative and non-perturbative interactions
between quarks via gluonic fields $A_\mu $. In accordance with the
approach [7], spin interactions can be treated as perturbations.
It is justified for $\rho$ and $K^*$ mesons. To construct the
expressions in spin interactions we write the relationship [7]
$$
\left\langle \exp \left\{ ig\Sigma _{\mu \nu }\left[
\int_0^sF_{\mu \nu }\left( z\right) dt-\int_0^{\bar s}F_{\mu \nu
}\left( \bar z\right) d\bar t\right] \right\} W(C) \right\rangle
$$
$$
=\exp \left\{ \Sigma _{\mu \nu }\left[ \int_0^sdt\frac \delta
{\delta \sigma _{\mu \nu }\left( t\right)
}-\int_0^{\overline{s}}d\bar t\frac \delta {\delta \sigma _{\mu
\nu }\left( \bar t\right) }\right] \right\}\left\langle
W(C)\right\rangle \eqno(2.6)
$$

where $\delta \sigma _{\mu \nu }\left( t\right) $ is the surface
around the point $z_\mu \left( t\right)$. The zeroth order in
spin-orbit and spin-spin interactions corresponds to neglecting
the terms $g\Sigma _{\mu \nu }F_{\mu \nu }$ in Eq. (2.4). We
suppose that mesons consist of quarks which move slowly with
respect to the time fluctuations of the gluonic fields ($T_q\gg
T_g$). It is the potential regime of the string. Voloshin [19] and
Leutwyler [20] remarked that in another case ($T_q\ll T_g$) the
dynamics is nonpotential and the QCD sum rules can be used. We
consider the case when the distance between quarks $r\gg T_g$.
Monte-Carlo calculations [21,22] gave $ T_g\simeq 0.2\div 0.3$ fm.
So we imply that the characteristic quark relative distance is
$r\simeq 1$ fm. This assumption will be confirmed below by the
calculation of the quark-antiquark relative coordinate.

The average Wilson integral (2.5) at large distances in accordance
with the area law can be represented in Minkowski space as
$$
\langle W(C) \rangle =\exp(i\sigma _0S), \eqno(2.7)
$$

where $\sigma _0$ is the string tension and $S$ is the area of the
minimal surface inside of the contour $C$. The surface $S$ can be
parametrized by the Nambu-Goto form [23,24]
$$
S=\int_0^Td\tau \int_0^1d\beta \sqrt{(\dot w_\mu w_\mu ^{\prime
})^2 -\dot w_\mu ^2w^{\prime }{}_\nu ^2}, \eqno(2.8)
$$

where $\dot w_\mu =\partial w_\mu /\partial \tau $, $w_\mu
^{\prime }=\partial w_\mu /\partial \beta $. Using the
approximation [7] that the coordinates of the string world surface
$w_\mu (\tau ,\beta )$ can be taken as straight lines for the
minimal surface we write
$$
w_\mu (\tau ,\beta )=z_\mu (\tau )\beta +\bar z_\mu (\tau
)(1-\beta ), \eqno(2.9)
$$

where $\tau $ is implied to be the proper time parameter for both
trajectories $\tau =(tT)/s=(\bar tT)/\bar s$. For uniform static
external electromagnetic fields we have the representation of the
vector potential through the strength tensor $F_{\mu \nu }^{el}$
$$
A_\nu ^{el}(z)=\frac 12F_{\mu \nu }^{el}z_\mu ,\qquad A_\nu
^{el}(\bar z) =\frac 12F_{\mu \nu }^{el}\bar z_\mu. \eqno(2.10)
$$

The paths $z_\mu $, $\bar z_\mu $ are expressed via the center of
mass coordinate $R_\mu $ and the relative coordinate $r_\mu $ [7],
$$
\bar z_\mu (\tau )=R_\mu -\frac{\bar s}{s+\bar s}r_\mu ,\qquad
z_\mu (\tau )=R_\mu +\frac s{s+\bar s}r_\mu \eqno(2.11)
$$

with the boundary conditions for $R_\mu (\tau )$, $r_\mu (\tau )$:
$$
R_\mu (0)=\frac{\mu _1y_\mu +\mu _2\bar y_\mu }{\mu _1+\mu
_2},\hspace{0.3in}R_\mu (T)= \frac{\mu _1x_\mu +\mu _2\bar x_\mu
}{\mu _1+\mu _2},
$$
$$
r_\mu (0)=y_\mu -\bar y_\mu ,\hspace{0.3in}r_\mu (T)=x_\mu -\bar
x_\mu.
$$

The integration with respect to $z_\mu $, $\bar z_\mu $ in (2.4)
is replaced by an integration over new variables $R_\mu $, $r_\mu
$. As $\tau $ is a common time for the quark and antiquark (the
time of the meson) the parametrization $z_\mu =(\tau ,{\bf z})$,
$\bar z_\mu =(\tau ,{\bf \bar z})$ is possible [7]. This leads to
the constraints: $R_0(\tau )=\tau $ , $ r_0(\tau )=0$. In
accordance with the approach in [7] we introduce the dynamical
masses $\mu _1$, $\mu _2$ by relationships
$$
\mu _1=\frac T{2s},\qquad \mu _2=\frac T{2\bar s}. \eqno(2.12)
$$

Replacing the integration with respect to $s$, $\bar s$ in Eq.
(2.4) by the integration over $d\mu _1$ and $d\mu _2$ with the
help of (2.7) - (2.11) we find [12] the two-point function in
zeroth order in the spin interactions
$$
G(x,\bar x;y,\bar y)=-T^2\int_0^\infty \frac{d\mu _1}{2\mu _1^2}
\int_0^\infty \frac{d\mu _2}{2\mu _2^2}\int DRDr\;exp\left\{
iS_{eff}\right\} \eqno(2.13)
$$

with the effective action
$$
S_{eff}=\int_0^Td\tau \biggl [-\frac{m_1^2}{2\mu
_1}-\frac{m_2^2}{2\mu _2} +\frac 12\left( \mu _1+\mu _2\right)
\dot R_\mu ^2+\frac 12\tilde \mu \dot r_\mu ^2
$$
$$
+\frac 12F_{\nu \mu }^{el}e\left( \dot R_\mu R_\nu +\frac 14\dot
r_\mu r_\nu \right) -\frac q4F_{\nu \mu }^{el}\left( \dot R_\mu
r_\nu +\dot r_\mu R_\nu \right)
$$
$$
-\int_0^1d\beta \sigma
_0\sqrt{(\dot w_\mu w_\mu ^{\prime })^2-\dot w_\mu ^2w^{\prime
}{}_\nu ^2}\biggr ], \eqno(2.14)
$$

where
$$
w_\mu =R_\mu +[\beta -\mu _1/(\mu _1+\mu _2)]r_\mu ;
\hspace{0.3in}\tilde \mu =\mu _1\mu _2/(\mu _1+\mu _2)
$$
is the reduced mass of the quark-antiquark system, $e=e_1+e_2$,
$q=e_1-e_2$. As a first step we are interested here in the
spinless part and therefore the preexponential terms $\left(
m_1+\frac i2\gamma _\mu \dot z_\mu \left( t\right) \right) $,
$\left( m_2-\frac i2\gamma _\mu \dot {\bar z}_\mu \left( \bar
t\right) \right) $ and the constant matrix $\ \Sigma _{\mu \nu
}F_{\mu \nu }^{el}$ was omitted in (2.13). The expression (2.14)
defines the effective Lagrangian for light mesons in external
uniform static electromagnetic fields in accordance with the
formula $S_{eff}=\int_0^T{\cal L} _{eff}d\tau $. The expression
(2.14) looks like a nonrelativistic one at $F_{\mu \nu }^{el}=0$,
but it is not. The author of [7] showed that the relativism is
contained here due to the $\tilde \mu $ dependence and the
spectrum is similar to that of the relativistic quark model.

The mass of the lowest states can be found on the basis of the
relationship [25]
$$
\int DRDr\exp\left\{ iS_{eff}\right\}
$$
$$
=\left\langle R=\frac{\mu_1x+\mu_2\bar x}{\mu_1+\mu_2},r=x-\bar
x\mid \exp\{-iT{\cal M}(\mu_1,\mu_2)\} \mid R=\frac{\mu _1y+\mu
_2\bar y}{\mu _1+\mu _2} ,r=y-\bar y\right\rangle, \eqno(2.15)
$$

where the mass ${\cal M}(\mu _1,\mu _2)$ is the eigenfunction of
the Hamiltonian. After that the Green function (2.13) is derived
by integrating Eq. (2.14) over the dynamical masses $\mu _1$, $\mu
_2$. In accordance with [7] we estimate the last integration on
$d\mu _1$, $d\mu _2$ using the steepest descent method which gives
a good accuracy when the Minkowski time $ T\rightarrow \infty $.
To have the correct formulas, it is necessary to go into Euclidean
space and return into Minkowski space on completing the functional
integration. We use this procedure.

\section{Ground State and Charge Radii}

The last term in (2.14) can be approximated with the accuracy of
$\sim 5$ \% [7] by the relation
$$
\int_0^1d\beta \sqrt{(\dot w_\mu w_\mu ^{\prime })^2-\dot w_\mu
^2w^{\prime }{}_\nu ^2}
$$
$$
=\int_0^1d\beta \sqrt{{\bf r}^2-\left( \beta -\frac{\mu _1}{ \mu
_1+\mu _2}\right) ^2(\dot {{\bf r}}\times {\bf r})^2}\simeq
\sqrt{{\bf r}^2} . \eqno(3.1)
$$

It is the potential regime at low orbital excitations of the
string when the orbital quantum number $l$ is small. As the
equalities $R_0(\tau )=\tau $, $ r_0(\tau )=0$ are valid, only
three-dimensional quantities are dynamical. Taking into account
(3.1), from Eq. (2.14) using the standard procedure we find the
canonical three-momenta corresponding to the center of mass
coordinate $R_\mu $ and the relative coordinate $r_\mu $
$$
\Pi_k=\frac{\partial {\cal L}_{eff}}{\partial \dot
{R_k}}=(\mu_1+\mu_2) \dot {R_k}+\frac e2F_{\nu k}^{el}R_\nu +\frac
q4F_{\nu k}^{el}r_\nu ,
$$
$$
\pi _k=\frac{\partial {\cal L}_{eff}}{\partial \dot {r_k}}=\tilde
\mu \dot {r_k}+\frac e8F_{\nu k}^{el}r_\nu +\frac q4F_{\nu
k}^{el}R_\nu. \eqno(3.2)
$$

The Hamiltonian
$$
{\cal H}=\pi _k\dot r_k+\Pi _k\dot R_k-{\cal L}_{eff}
$$

found from Eq. (2.14) with the help of Eqs. (3.1), (3.2) takes the
form
$$
{\cal
H}=\frac{m_1^2}{2\mu_1}+\frac{m_2^2}{2\mu_2}+\frac{\mu_1+\mu_2}
{2}+ \frac{ \mu_1+\mu_2}{2}\dot R_k^2+\frac{\tilde \mu }2\dot
r_k^2-\frac e2( {\bf E} {\bf R})
$$
$$
-\frac q4({\bf E}{\bf r})+\sigma_0\sqrt{{\bf r}^2} \eqno(3.3)
$$

so that the equation for the eigenvalues is given by
$$
{\cal H}\Phi ={\cal M}(\mu_1,\mu_2)\Phi. \eqno(3.4)
$$

The terms containing the strength of the electric field in Eq.
(3.3) describe the interaction of the dipole electric moment ${\bf
d}$ with an external electric field. Using the definitions we have
$$
\frac e2({\bf E}{\bf R})+\frac q4({\bf E}{\bf r})=\frac 12(e_1 {\bf r_1}+e_2
{\bf r_2}){\bf E}={\bf d}{\bf E} \eqno(3.5)
$$

and the interaction energy of the electric dipole moment with a
uniform static electric field is $U=-{\bf d}{\bf E}$. Bellow we
investigate the case of a pure magnetic field when ${\bf E}=0$.
The case ${\bf E}\neq 0$, ${\bf H}=0$ was considered in [12].
Using Eq. (3.2) the equation for the eigenfunction $\Phi $ of the
auxiliary ``Hamiltonian"
$$
\tilde {{\cal H}}={\cal H} -m_1^2/2\mu _1-m_2^2/2\mu _2-(\mu
_1+\mu _2)/2
$$

is given by
$$
\biggl [\frac 1{2\tilde \mu }\left({\bf \pi}-\frac e8({\bf
r}\times {\bf H} )-\frac q4({\bf R}\times {\bf H})\right)^2
$$
$$
+\frac 1{2(\mu _1+\mu _2)}\left({\bf \Pi}-\frac e2({\bf R}\times
{\bf H} )-\frac q4({\bf r}\times {\bf H})\right)^2+\sigma
_0\sqrt{{\bf r}^2} \biggr ] \Phi
$$
$$
=\epsilon (\mu ,{\bf H})\Phi ,\eqno(3.6)
$$

where $\epsilon (\mu ,{\bf H})$ is the eigenvalue. In accordance
with the Noether theorem we come to the conclusion that the
canonical momentum ${\bf \Pi}$ corresponding to the center of mass
coordinate is a constant, i.e., $ {\bf \Pi }=const$. Therefore it
is possible to choose the condition ${\bf \Pi }=0$. Putting ${\bf
\Pi }=0$ and ${\bf R}=0$ into Eq. (3.6) we arrive to the equation
$$
\left [ \frac 1{2\tilde \mu }\left({\bf \pi}-\frac e8({\bf
r}\times {\bf H} )\right)^2+\frac{q^2}{32(\mu _1+\mu _2)}({\bf
r}\times {\bf H})^2+\sigma _0 \sqrt{{\bf r}^2}\right ] \Phi
$$
$$
=\epsilon (\mu ,{\bf H} )\Phi . \eqno(3.7)
$$

The second term in (3.7) describes the effect of the recoil of the
string. Such a term appears also in non-relativistic models
[26,13,14]. If we put ${\bf R}=0$ in Eq. (2.13), this term would
not appear [12].

In quantum theory instead of the path integration in ${\bf r}$ we
can use the replacement $\pi_{k} \rightarrow - i\partial/\partial
r_{k}$. We can apply Eq. (3.7) to the leading trajectories with
light quarks with masses $m_1=m_2 \equiv m$, $\mu_1=\mu_2 \equiv
\mu$ ($\tilde {\mu} =\mu/2$) for $\rho$ meson and when $m_1\neq
m_2$, $\mu_1\neq \mu_2$ for $K^*$ meson.

An external magnetic field splits the energy levels like the
Zeeman effect for atoms. The difference with our case is we
describe here the light quark-antiquark system in center mass
system (c.m.s.) with the linear potential between quarks.
Therefore the spectrum of the energy has other levels.

We can consider a small external magnetic field so that here
perturbative theory can be applied. We receive a first
approximation when the external field ${\bf H}$ is switched off
(${\bf H}=0$) and the equation for the eigenvalue is given by
$$
\left( -\frac 1{2\tilde \mu }\frac{\partial ^2}{\partial r_i^2}+\sigma _0
\sqrt{{\bf r}^2}\right) \Phi =\epsilon (\mu )\Phi . \eqno(3.8)
$$

Equation (3.8) gives the discrete values of the energy $\epsilon
(\mu )$ due to the shape of the potential energy. The numerical
solution of equation (3.8) was obtained in [27]. It is useful to
find the solution to Eq. (3.8) for the ground state in an
analytical form. After introducing the variables $\rho _k=(2\tilde
\mu \sigma _0)^{1/3}r_k$, $\epsilon (\tilde \mu )=(2\tilde \mu
)^{-1/3}\sigma _0^{2/3}a(n)$ [7], Eq. (3.8) becomes
$$
\left( -\frac{\partial ^2}{\partial \rho _i^2}+\rho \right) \Phi (\rho
)=a(n)\Phi (\rho ). \eqno(3.9)
$$

The solution to Eq. (3.9) may be chosen in the form $\Phi (\rho
)=R(\rho )Y_{lm}(\theta ,\phi )$, where $Y_{lm}(\theta ,\phi )$
are spherical functions. After setting the variable $R(\rho )=\chi
/\rho$ we come to the equation for the radial function
$$
\chi
^{^{\prime \prime }}(\rho )+\left( a(n)-\rho -\frac{l(l+1)}{\rho
^2} \right) \chi (\rho )=0, \eqno(3.10)
$$

where $\chi ^{^{\prime \prime }}(\rho )=\partial ^2\chi (\rho
)/\partial \rho ^2$, and $l$ is an orbital quantum number. The
solutions to Eq. (3.10) for the ground state $l=0$ are the Airy
functions $Ai(\rho -a(n))$, $Bi(\rho -a(n))$ [28]. The finite
solution to Eq. (3.10) at $\rho \rightarrow \infty $ $ (l=0)$ is
$$
\chi (\rho )=NAi(\rho -a(n)). \eqno(3.11)
$$

The constant N can be found from the normalization condition
$$
\int_0^\infty \chi ^2(\rho )d\rho =1. \eqno(3.12)
$$

The requirement that this solution satisfies the condition
$$
\chi (0)=NAi(-a(n))=0
$$

gives the Airy function zeroes [28],
$$
a(1)\equiv a_1=2.3381 ,\hspace{0.3in}a(2)\equiv a_2=4.0879
$$

and so on. The principal quantum number $n=n_r+l+1$, where $n_r$
is the radial quantum number which defines the number of zeroes of
the function $\chi (\rho )$ at $\rho >0$. For the ground state we
should take the solution (3.11) at $a(n)=a_1$~ (here $n_r=0$,
$l=0$):
$$
\chi _0(\rho )=N_0Ai(\rho -a_1). \eqno(3.13)
$$

Now let us estimate the mean-squared radius for the state which is
described by the function $\Phi $ [the solution of equation
(3.8)]. Multiplying Eq. (3.8) by the conjugated function $\Phi
^{*}$ and integrating over the volume we find the relations
$$
\left\langle T\right\rangle +\left\langle U\right\rangle =\epsilon (\tilde
\mu ),
$$
$$
\left\langle T\right\rangle =-\frac 1\mu \int \Phi ^{*}\partial
_k^2\Phi dV,\hspace{0.3in}\left\langle U\right\rangle =\sigma
_0\int \sqrt{{\bf r}^2}\Phi ^{*}\Phi dV. \eqno(3.14)
$$

It is seen from Eqs. (3.14) that the mean potential energy
$\left\langle U\right\rangle =\sigma _0\left\langle \sqrt{{\bf
r}^2}\right\rangle $ is connected to the mean diameter
$\left\langle \sqrt{{\bf r}^2} \right\rangle $ ~(because ${\bf r}$
is the relative coordinate and quarks move around their center
mass), which defines the size of mesons. In accordance with the
virial theorem [29] we have the connection of the mean kinetic
energy with the mean potential energy
$$
2\left\langle T\right\rangle =k\left\langle U\right\rangle, \eqno(3.15)
$$

where $k$ is defined from the equality $U(\lambda r)=\lambda
^kU(r)$. In our case of the linear potential $k=1$ and from Eqs.
(3.14), (3.15) we get
$$
\left\langle U\right\rangle =\frac 23\epsilon (\tilde \mu )=\frac
23 (2\tilde \mu )^{-1/3}\sigma _0^{2/3}a(n). \eqno(3.16)
$$

The use of the steepest descent method for the estimation of the
integration in $\mu $~ (at ${\bf H}=0$) leads to the conditions
[7]:
$$
\frac{\partial {\cal M}(\mu _1,\mu _2)}{\partial \mu
_1}=0,\hspace{0.3in}\frac{
\partial {\cal M}(\mu _1,\mu _2)}{\partial \mu _2}=0, \eqno(3.17)
$$

where the mass of the ground state ${\cal M}(\mu _1,\mu _2)$ is
given by [see Eqs. (3.3), (3.4)]
$$
{\cal M}(\mu_1,\mu_2)=\frac{m_1^2}{2\mu_1}+\frac{m_2^2}{2\mu_2}+\frac{
\mu_1+\mu_2} 2+(2\tilde \mu )^{-1/3}\sigma_0^{2/3}a(n). \eqno(3.18)
$$

Here we consider the more general case as compared with [7] when
$\mu _1\neq \mu _2$~ ($m_1\neq m_2$). This case is realized for
$K^*$ mesons. It is assumed that the current mass of u,d quarks
($m_u=5.6\pm 1.1 $ MeV, $ m_d=9.9\pm 1.1$ MeV [30]), $m_1$ is much
less than the dynamical mass $\mu _1$ ($\mu _1\simeq 330$ MeV),
i.e., $m_1\ll \mu _1$ and the mass of s quarks $m_2$ ~($m_s=199\pm
33$ MeV [30]), is comparable with $\mu _1$ but $m_2<\mu _1$. Using
these assumptions we neglect the term $m_1^2/2\mu _1$ in Eq.
(3.18) and from Eqs. (3.17) have the equations
$$
(2\tilde \mu \sigma _0)^{2/3}a(n)=3\mu
_1^2,\hspace{0.3in}3m_2^2+(2\tilde \mu \sigma _0)^{2/3}a(n)=3\mu
_2^2. \eqno(3.19)
$$

From Eqs. (3.19) we arrive to the expression for the dynamical
mass $\mu _2$ (for s quarks):
$$
\mu _2=\sqrt{\mu _1^2+m_2^2}. \eqno(3.20)
$$

To find $\mu _1$ the perturbation in the parameter $m_2^2/\mu
_1^2$ will be assumed. Using the relation $\mu _2\simeq \mu
_1(1+m_2^2/(2\mu _1^2))$ which is obtained from Eq. (3.20) and the
definition of the reduced mass $\tilde \mu =\mu _1\mu _2/(\mu
_1+\mu _2)$ from Eqs. (3.19) we arrive at the equation
$$
\mu _1\simeq \sqrt{\sigma _0}\left( \frac{a(n)}3\right) ^{3/4}
\left( 1+\frac{ m_2^2}{8\mu _1^2}\right). \eqno(3.21)
$$

In the zeroth order we come to the value $\mu _0\equiv \mu _1^{(0)}=\sqrt{
\sigma _0}(a(n)/3)^{3/4}$ [7]. The next order gives the relationship
$$
\mu _1\simeq \sqrt{\sigma _0}\left( \frac{a(n)}3\right) ^{3/4}+
\frac{m_2^2}{ 8 \sqrt{\sigma _0}}\left( \frac 3{a(n)}\right)
^{3/4}. \eqno(3.22)
$$

In a particular case $m_2=0$ we arrive at $\mu _1=\mu _2=\mu
_0=\sqrt{\sigma _0}(a(n)/3)^{3/4}$ [7]. The value of the string
tension $\sigma _0=0.15$ $\mbox{GeV}^2$ was found from a
comparison of the experimental slope of the linear Regge
trajectories $\alpha ^{\prime }=0.85$ $\mbox{GeV}^{-2}$, and the
variable $\alpha ^{\prime }=1/8\sigma _0$ [7]. It leads for the
lowest state $n_r=0,l=0,a(1)=2.3381$ to the value $\mu _0=321$ MeV
[7]. This means that for $\rho $ mesons when $m_1=m_u$, $m_2=m_d$
we have the dynamical masses of $u$, $d$ quarks $\mu _1=\mu _2=\mu
_0$. For $K^*$ mesons using Eq. (3.20) and $ m_2=m_s\simeq 200$
MeV [30] from Eq. (3.22) we get the reasonable values
$$
\mu _1\simeq 337~\mbox{MeV},\hspace{0.3in}\mu _2\simeq
392~\mbox{MeV},\hspace{0.3in}\tilde \mu \simeq 181~\mbox{MeV} .
\eqno(3.23)
$$

Inserting the equation $\left\langle U\right\rangle =\left\langle
\sigma _0\sqrt{ {\bf r}^2}\right\rangle $ into the left-hand side
of Eq. (3.16) one gives the expression
$$
\left\langle \sqrt{{\bf r}^2}\right\rangle =\frac 23(2\tilde \mu
\sigma _0)^{-1/3}a(n). \eqno(3.24)
$$

From Eqs. (3.20), (3.22) using the first order in the parameter
$m_2^2/\mu _1^2$ we find
$$
2\tilde \mu \simeq \mu _0\left( 1+\frac{3m_2^2}{8\mu _0^2}\right)
\hspace{0.3in}\left[ \mu _0=\sqrt{\sigma _0}\left(
\frac{a(n)}3\right) ^{3/4}\right] . \eqno(3.25)
$$

Equation (3.24) with the help of Eq. (3.25) gives an approximate
relation for the mean relative coordinate:
$$
\left\langle \sqrt{{\bf r}^2}\right\rangle =\frac 2{\sqrt{\sigma
_0} } \left( \frac{a(n)}3\right) ^{3/4}\left[
1+\frac{3m_2^2}{8\sigma _0}\left( \frac 3{a(n)}\right)
^{3/2}\right] ^{-1/3}. \eqno(3.26)
$$

For $\rho$ mesons putting $m_2=0$ in Eq. (3.26) we arrive at
$$
\left\langle \sqrt{{\bf r}^2}\right\rangle =\frac 2{\sqrt{\sigma
_0} } \left( \frac{a(n)}3\right) ^{3/4}. \eqno(3.27)
$$

The same expression (3.27) was found in [12] using another method.
With the help of the definition of the center of mass coordinate
we can write an approximate relation for the mean charge radius of
$\rho$-mesons: \footnote{ The relationship $\sqrt{\left\langle
{\bf r}^2\right\rangle }\simeq \left\langle \sqrt{{\bf
r}^2}\right\rangle $ is confirmed by the numerical calculations.}
$$
\sqrt{\left\langle r_{\rho}^2\right\rangle }\simeq \frac{1}{2}
\left\langle \sqrt{{\bf r}^2}\right\rangle. \eqno(3.28)
$$

At $\sigma _0=0.15$ $\mbox{GeV}^2$ [7,11] and $a(1)=2.3381$
Eqs.(3.27), (3.28) give
$$
\sqrt{\left\langle r_{\rho}^2\right\rangle }\simeq
0.42~\mbox{fm}\hspace{0.3in}\left( \left\langle \sqrt{{\bf
r}^2}\right\rangle =0.84~\mbox{fm}\right) . \eqno(3.29)
$$

The value (3.29) characterizes the radius of the sphere where the
wave function of the $\rho$ meson is concentrated (remember that
${\bf r}$ is the distance between quarks). We know only the
experimental data for $\pi^{\pm}$ mesons which have the same quark
structure as $\rho^{\pm}$ mesons:
$$
\left\langle r_{\pi ^{\pm }}^2\right\rangle_{exp}=(0.44\pm
0.02)~\mbox{fm}^2\hspace{0.3in}\left( \sqrt{ \left\langle r_{\pi
^{\pm }}^2\right\rangle }_{exp}\simeq
0.66~\mbox{fm}\right)\hspace{0.3in} [31].
$$

For calculating the relative coordinate of $K^*$ mesons we should
use Eq. (3.24) or (3.26) with the conditions $\mu _1=\mu _u$ and
$\mu _2=\mu _s$ Eqs. (3.23). As a result formula (3.24) gives the
value of the mean relative coordinate of $K^*$ mesons:
$$
\left\langle \sqrt{{\bf r}^2}\right\rangle _{K^*}=0.79~\mbox{fm}.
\eqno(3.30)
$$

With the help of this relation we can estimate the mean charge
radius of $ K^* $ mesons:
$$
\sqrt{\left\langle r_{K^*}^2\right\rangle }\simeq \frac{\mu
_2}{\mu _1+\mu _2 } \left\langle \sqrt{{\bf r}^2}\right\rangle
_{K^*}=0.54\left\langle \sqrt{{\bf r}^2}\right\rangle
_{K^*}=0.43~\mbox{fm}. \eqno(3.31)
$$

The experimental data of the mean charge radius of $K^{\pm }$
mesons having an analogous quark structure as $K^*$ mesons are
$$
\sqrt{\left\langle r_{K^{\pm }}^2\right\rangle }=(0.53\pm
0.05)~\mbox{fm} \hspace{0.3in} [32],
$$
$$
\left\langle r_{K^{\pm }}^2\right\rangle =(0.34\pm
0.05)~\mbox{fm}^2\hspace{0.3in} [31]
$$

and for neutral $K^0$ mesons
$$
\sqrt{\left\langle r_{K^0}^2\right\rangle }=(0.28\pm 0.09)
\mbox{fm} \hspace{0.3in}[32].
$$

 So expression (3.26) gives a reasonable value for the
charge radius of $ K^*$ mesons, although we need the experimental
data of the charge radius of $ K^{*}$ mesons .

The first perturbative one-gluon exchange contribution to
Hamiltonian (3.3) determines the spin-spin correction such as the
Breit-Fermi hyperfine interaction [9]. The spin-spin interaction
is important to explain the Nambu-Goldstone phenomenon which takes
place for the $l=s=0$ channel.

\section{Perturbative Expansion and Magnetic
Polarizabilities}

To calculate the magnetic polarizability of mesons in accordance
with Eq. (1.1) we should know the Hamiltonian depending on the
magnetic field ${\bf H}$. From Eq. (3.7) we arrive at the
expression of the auxiliary ``Hamiltonian":
$$
\tilde {{\cal H}}=-\frac 1{2\tilde \mu }\frac{\partial
^2}{\partial r_k^2} +\frac e{8\tilde \mu }{\bf H}{\bf L}+\left(
\frac{e^2}{128\tilde{\mu}}+ \frac{q^2}{ 32(\mu _1+\mu _2)}\right)
\left[ {\bf r}^2{\bf H}^2-( {\bf r} {\bf H} )^2\right]
$$
$$
 +\sigma_0\sqrt{{\bf r}^2}, \eqno(4.1)
$$

where ${\bf L}=-i({\bf r}\times {\bf \partial})$~ ($\partial _k=\partial
/\partial r_k$) is the angular momentum. Considering the external magnetic
field ${\bf H}=(0,0,H)$ expression (4.1) is rewritten as
$$
{\tilde {{\cal H}}}=-\frac 1{2\tilde \mu }\frac{\partial
^2}{\partial r_i^2}+ \frac{ eH}{8\tilde \mu }L_3+\left(
\frac{e^2}{4\tilde \mu }+\frac{q^2}{ \mu _1+\mu _2}\right)
\frac{H^2}{32}\left( r_1^2+r_2^2\right) +\sigma _0 \sqrt{ {\bf
r}^2 }, \eqno(4.2)
$$

where $L_3=i(r_2\partial _1-r_1\partial _2)$ is the third
projection of the angular momentum.

Now let us consider the magnetic polarizability of mesons on the
basis of perturbative theory. We can rewrite Eq. (4.2) in the form
$$
\tilde {{\cal H}}={\cal H}_0+\frac{eH}{8\tilde \mu }L_3+\left(
\frac{ e^2}{ 4\tilde \mu }+ \frac{q^2}{\mu_1+\mu_2}\right)
\frac{H^2r^2\sin^2\theta }{32}, \eqno(4.3)
$$

where $\theta $ is the angle between ${\bf H}$ and the relative
coordinate $ {\bf r}$ and the free Hamiltonian is given by
$$
{\cal H}_0=-\frac 1{2\tilde \mu }\frac{\partial ^2}{\partial
r_k^2} +\sigma_0 \sqrt{{\bf r}^2}.
$$

For the small magnetic fields ${\bf H}$ the second and third terms
of Eq. (4.3) can be considered as a perturbation. Then using the
standard perturbative method [33], we find a shift of the energy
in the state $\mid n\rangle$ with the accuracy of second order in
${\bf H}$:
$$
\Delta E_n=\langle n\mid \frac{eH}{8\tilde \mu }L_3+\left(
\frac{e^2}{ 4\tilde \mu }+\frac{q^2}{\mu _1+\mu _2}\right)
\frac{H^2r^2\sin^2\theta }{32} \mid n\rangle
$$
$$
+\sum_{n^{^{\prime }}}\frac{\mid \langle n^{^{\prime }}\mid
eHL_3/(8\tilde \mu) \mid n\rangle \mid ^2}{E_n-E_{n^{^{\prime }}}}
\eqno(4.4)
$$

For the ground state $l=0$ ($s$ state) the first and third terms
of Eq. (4.4) do not give a contribution to the energy because
$L_3\mid 0\rangle =0$. Taking the mean value and using the
condition $(1/4\pi )\int \sin^2\theta d\Omega =2/3$, from Eq.
(4.4) we come to
$$
\Delta E_0=\left( \frac{e^2}{4\tilde \mu }+\frac{q^2}{\mu _1+\mu
_2} \right) \frac{H^2}{48}\left\langle {\bf r}^2\right\rangle .
\eqno(4.5)
$$

Comparing Eq. (4.5) with Eq. (1.1) we find the magnetic
polarizability of light mesons
$$
\beta =-\frac 1{24}\left( \frac{e^2}{4\tilde \mu }+\frac{q^2}{\mu
_1+ \mu _2} \right) \left\langle {\bf r}^2\right\rangle .
\eqno(4.6)
$$

It should be noticed that here $\left\langle {\bf
r}^2\right\rangle $ means the mean-squared relative coordinate of
the quark-antiquark system. Expression (4.6) is similar to the
Langevin formula for the magnetic susceptibility of atoms. It is
seen that we have here only the diamagnetic polarizability as
$\beta <0$. To calculate the paramagnetic polarizability one needs
to take into account the interaction of the meson spin with the
magnetic field. Using the approximate relation $\left\langle {\bf
r} ^2\right\rangle \simeq \left\langle \sqrt{{\bf r}^2}
\right\rangle ^2$, parameters $e=e_1+e_2$, $q=e_1-e_2$, expression
(3.27), and the dynamical masses of $u,d$ quarks $\mu _1=\mu _2$,
$\widetilde{\mu }=\mu /2$ we find from Eq. (4.6) the relationship
for the diamagnetic polarizability of $\rho$ mesons:
$$
\beta _\rho \simeq -\frac{(e_1^2+e_2^2)}{6\mu \sigma _0}\left(
\frac{a(n)} 3\right) ^{3/2}, \eqno(4.7)
$$

Calculating Eq. (4.7) for charged $\rho$ mesons at $\sigma_0=0.15$
$\mbox{GeV}^2$, $\mu =\mu _0=321$ GeV, $e_1=2e/3$, $e_2=e/3$,
$e^2=1/137$, $n=1$ in Gaussian units one takes the value
$$
\beta _{\rho}\simeq -0.8\times 10^{-4}~\mbox{fm}^3. \eqno(4.8)
$$

In rationalized units, the polarizability is $4\pi $ times greater.

For $K^*$ mesons, Eq. (4.6) at the values (3.23), (3.31) $\left(
\left\langle {\bf r}^2\right\rangle \simeq \left\langle \sqrt{{\bf
r}^2}\right\rangle ^2\right) $ leads to the magnitude
$$
\beta _{K^{*}}=-0.57\times 10^{-4}~\mbox{fm}^3. \eqno(4.9)
$$

Unfortunately there are not experimental data of $\rho$ and $K^*$
meson polarizabilities yet.

\section{Conclusion}

The QCD string theory allows us to estimate the mean squared radii
and magnetic polarizabilities of $\rho $, $K^*$ mesons. These
quantities were derived as functions of the string tension which
is a fundamental variable in this approach. It is not difficult to
calculate the magnetic polarizabilities of excited states of
mesons using this approach. For that we should take the quantum
numbers $n_r=1$, $l=0$ ~$(n=2)$ and evaluate the mean relative
coordinate in accordance with Eq. (3.26). Then Eq. (4.6) gives the
necessary polarizabilities. To have more precise values of the
meson electromagnetic characteristics one needs to take into
account spin corrections. Especially it is important for light
pseudoscalar mesons ($\pi$, $K$ mesons). In principle it is
possible to receive spin-orbit and spin-spin interactions using
the general expressions (2.4), (2.6).

The Nambu-Jona-Lasinio (NJL) model [34] having a good basis in the
framework of QCD [35] describes chiral symmetry breaking but not
the confinement of quarks [7,36]. Besides this model has free
parameters and the calculated polarizabilities of mesons [37] are
parameter dependent.

The instanton vacuum theory (IVT) developed in [39-40] does not
give the confinement of quarks phenomenon [7]. This theory is like
the NJL model [36] and takes into account only chiral symmetry
breaking. Therefore the calculation of the meson electromagnetic
polarizabilities on the basis of the IVT gave the similar result
[41] as in the NJL model.

All this shows that the theoretical evaluation of the charge radii
and the magnetic polarizabilities of $\rho$, $K^*$ mesons is
possible on the basis of a good description of chiral symmetry
breaking and the confinement of quarks in the framework of QCD
string theory but with some approximations and model assumptions.
Naturally that theory was derived using the nonperturbative QCD,
i.e., first principals of QCD.

\begin{center}
{\bf APPENDIX}
\end{center}

In this appendix we derive the one-quark Green function using the
Fock-Schwinger method. Starting with the approach [15] and
introducing external electromagnetic and gluonic fields we write
the Green function of quarks which possess spins in Minkowski
space
$$
S(x,y)=\left\langle x\mid \left( \widehat{D}+m_1\right)^{-1}\mid y
\right\rangle =\left\langle x\mid \left( m_1-\widehat{D}\right)
\left( m_1^2- \widehat{D}^2\right)^{-1}\mid y \right\rangle
$$
$$
=\left\langle \psi \left( x\right) \overline{\psi }\left( y\right)
\right\rangle , \eqno(A1)
$$

where $e_1$ and $m_1$ are the charge and mass of the quark,
$\widehat{D} =\gamma _\mu D_\mu $, $D_\mu =\partial _\mu
-ie_1A_\mu ^{el}-igA_\mu $, $ A_\mu =A_\mu ^a\lambda ^a$; $\gamma
_\mu $ and $\lambda ^a$ are the Dirac and Gell-Mann matrices,
respectively; $A_\mu ^{el}$ and $A_\mu ^a$ are the electromagnetic
and gluonic vector potentials, respectively. The inverse operator
$\left( m_1^2-\widehat{D}^2\right) ^{-1}$ can be represented in
the proper time $s$ [16]:
$$
\left( m_1^2-\widehat{D}^2\right) ^{-1}=i\int_0^\infty ds\exp \left\{
-is\left( m_1^2-\widehat{D}^2\right) \right\} . \eqno(A2)
$$

Using the properties of Dirac matrices $\left\{ \gamma _\mu
,\gamma _\nu \right\} =2\delta _{\mu \nu }$ we find the squared
operator
$$
\widehat{D}^2=D_\mu ^2+\Sigma _{\mu \nu }\left( e_1F_{\mu \nu
}^{el}+ gF_{\mu \nu }\right) , \eqno(A3)
$$

where
$$
\Sigma _{\mu \nu }=-\frac i4[\gamma _\mu ,\gamma _\nu
],\hspace{0.5in} F_{\mu \nu }^{el}=\partial _\mu A_\nu
^{el}-\partial _\nu A_\mu ^{el},
$$
$$
F_{\mu \nu }=\partial _\mu A_\nu -\partial _\nu A_\mu -ig[A_\mu ,A_\nu ],
$$

$\Sigma _{\mu \nu }$ are the spin matrices, and $F_{\mu \nu
}^{el}$, $F_{\mu \nu }$ are the strength of electromagnetic and
gluonic fields, respectively. Inserting relationship (A2) into Eq.
(A1) with the help of Eq. (A3) we get
$$
S(x,y)=i\int_0^\infty ds\langle x\mid \left(
m_1-\widehat{D}\right) \exp \biggl\{ -is\biggl[ m_1^2-D_\mu ^2
$$
$$
-\Sigma _{\mu \nu }\left( e_1F_{\mu \nu }^{el}+gF_{\mu \nu
}\right) \biggr] \biggr\} \mid y\rangle . \eqno(A4)
$$

The exponent in Eq. (A4) plays the role of the evolution operator
which defines the dynamics of the ``Hamiltonian" $m_1^2-D_\mu
^2-\Sigma _{\mu \nu }\left( e_1F_{\mu \nu }^{el}+gF_{\mu \nu
}\right) $ with initial $\mid y\rangle $ and final $\langle x\mid
$ states where $s$ means the proper time. Therefore it is
convenient to represent the matrix element in Eq. (A4) as a path
integral [16]:
$$
S(x,y)=i\int_0^\infty dsN\int_{z(0)=y}^{z(s)=x}DpDzP\left(
m_1-\widehat{D} \right) \exp \biggl [i\int_0^sdt\biggl [p_\mu \dot
z_\mu -m_1^2
$$
$$
-\left( p_\mu -e_1A_\mu ^{el}-gA_\mu \right) ^2+\Sigma _{\mu \nu
}\left( e_1F_{\mu \nu }^{el}+gF_{\mu \nu }\right) \biggr ] \biggr
], \eqno(A5)
$$

where $z_\mu \left( t\right) $ is the path of a quark with the
boundary conditions $z(0)=y$, $z(s)=x$, $\widehat{D}=i\gamma _\mu
\left( p_\mu -e_1A_\mu ^{el}-gA_\mu \right) $ and $P$ means
ordering; $N$ is a constant which is connected to the measure
definition and it will be chosen later. The path integration over
the momenta can be rewritten in the form (see [7,18]
$$
N\int Dp\left( m_1-\widehat{D}\right) \exp \left\{
i\int_0^sdt\left[ p_\mu \dot z_\mu -\left( p_\mu -e_1A_\mu
^{el}-gA_\mu \right) ^2\right] \right\}
$$
$$
=N\int Dp\exp \left[ i\int_0^sdt\left( p_\mu \dot z_\mu \right)
\right] \left( m_1+\frac 12\gamma _\mu \frac \delta {\delta p_\mu
}\right)
$$
$$
\times \exp \left\{ -i\int_0^sdt\left( p_\mu -e_1A_\mu
^{el}-gA_\mu \right) ^2\right\}
$$
$$
=N\int Dp\left( m_1-\frac i2\gamma _\mu \dot z_\mu \right) \exp
\left\{ i\int_0^sdt\left[ p_\mu \dot z_\mu -\left( p_\mu -e_1A_\mu
^{el}-gA_\mu \right) ^2\right] \right\}
$$
$$
=N\int Dp\left( m_1-\frac i2\gamma _\mu \dot z_\mu \right) \exp
\left\{ i\int_0^sdt\left[ -p_\mu ^2+\frac 14\dot z_\mu ^2+\left(
e_1A_\mu ^{el}+gA_\mu \right) \dot z_\mu \right] \right\} .
\eqno(A6)
$$

In Eq. (A6) we used the integration by parts (see [18]) and made a
continuity of shifts $p_\mu \rightarrow p_\mu +e_1A_\mu
^{el}+gA_\mu $ and then $p_\mu \rightarrow p_\mu +\dot z_\mu /2$.
The constant $N$ in Eq. (A6) is defined by the relation
$$
N\int Dp\exp \left\{ -i\int_0^sdt\left( p_\mu ^2\right) \right\}
=1. \eqno(A7)
$$

Taking into account Eqs. (A6), (A7) we find from Eq. (A5) the
Green function of the quark
$$
S(x,y)=i\int_0^\infty ds\int_{z(0)=y}^{z(s)=x}Dz\;\left( m_1-\frac
i2\gamma _\mu \dot z_\mu \left( t\right) \right) P\exp
{}{}{}\biggl \{i\int_0^sdt \biggl [\frac 14\dot z_\mu ^2\left(
t\right) -m_1^2
$$
$$
+\left( e_1A_\mu ^{el}+gA_\mu \right) \dot z_\mu \left( t\right)
+\Sigma _{\mu \nu }\left( e_1F_{\mu \nu }^{el}+gF_{\mu \nu
}\right) \biggr ] \biggr \}. \eqno(A8)
$$

\end{document}